\documentclass[english,aps,superscriptaddress,showkeys,showpacs,nofootinbib]{revtex4}

\usepackage{booktabs}
\usepackage{siunitx}

\pdfoutput=1
\usepackage[T1]{fontenc}
\usepackage[letterpaper]{geometry}
\usepackage{units}
\usepackage{bbding}
\usepackage{amsmath}
\usepackage{amssymb}
\usepackage{fixmath}
\usepackage{graphicx}
\usepackage{esint}
\usepackage{tensor}
\usepackage[colorinlistoftodos,prependcaption,textsize=tiny]{todonotes}

\usepackage[utf8]{inputenc}
\usepackage{lmodern} 

\makeatletter

\newcolumntype{C}{>{$}c<{$}}
\newcolumntype{L}{>{$}l<{$}}
\newcolumntype{R}{>{$}r<{$}}
\newcommand\Tstrut{\rule{0pt}{4.6ex}}       
\newcommand\Bstrut{\rule[-1.2ex]{0pt}{0pt}} 

\AtBeginDocument{
\heavyrulewidth=.08em
\lightrulewidth=.05em
\cmidrulewidth=.03em
\belowrulesep=.65ex
\belowbottomsep=0pt
\aboverulesep=.4ex
\abovetopsep=0pt
\cmidrulesep=\doublerulesep
\cmidrulekern=.5em
\defaultaddspace=.5em
}

\usepackage{doi}
\usepackage{hyperref}

\makeatother

\usepackage{babel}

  \def\refeq#1{Eq.~(\ref{#1})}
  \def\refeqs#1{Eqs.~(\ref{#1})}
  \def\refeqa#1{~(\ref{#1})}

  \def\refsec#1{Section\ \ref{#1}}

  \newenvironment{eqaligned}
    {\begin{equation}
    \begin{aligned}
    }
    {
    \end{aligned}
    \end{equation}
    }
  \newcommand{\vect}[1]{ \mathbold{#1}}  
  \newcommand{\defn}{ \equiv}

  \newcommand{\lp}{\left(}
  \newcommand{\rp}{\right)}
  \newcommand{\lb}{\left[}
  \newcommand{\rb}{\right]}
  \newcommand{\la}{\left<}
  \newcommand{\ra}{\right>}

  \newcommand{\vB}{\vect{B}}
  \newcommand{\vJ}{\vect{J}}
  \newcommand{\vA}{\vect{A}}
  
  \newcommand{\vV}{\vect{V}}
  \newcommand{\vF}{ \vect{F} }

  \newcommand{\specheat}{\gamma_h}

  \newcommand{\grad}{\vect{\nabla}}
  \newcommand{\curl}[1]{\grad \times #1 }
  \newcommand{\dive}[1]{\grad \cdot #1 }

  \newcommand{\dpsi}[1]{\frac{\partial #1}{\partial \psi}}

  \newcommand{\jac}{{\mathcal{J}}}
  \newcommand{\jaci}{{\mathcal{J}}^{-1}}
  \newcommand{\Pp}{ P^\prime }				
  \newcommand{\Vp}{V^\prime}
  \newcommand{\Vpp}{V^{\prime\prime}}
  \newcommand{\Vpo}{ \frac{V^\prime}{4 \pi^2}}
  \newcommand{\norm}{ P^\prime }		
  \newcommand{\RR}{ \psi }			
  \newcommand{\vR}{ \grad \RR }		
  \newcommand{\C}{ C }				
  \newcommand{\vC}{ \vect{\C} }		
  \newcommand{\vK}{ \vect{K} }			
  \newcommand{\vRsq}{ \mid \grad \RR \mid^2 }
  \newcommand{\vCsq}{ \C^2 }
  \newcommand{\vKsq}{ K^2 }
  \newcommand{\vBsq}{ B^2 }
  \newcommand{\vrr}{\frac{ \vR}{\vRsq} }
  \newcommand{\vbb}{\frac{ \vB}{B^2} }
  \newcommand{\vcc}{\frac{ \vC}{\vCsq} }
  \newcommand{\vjj}{\frac{ \vJ}{J^2} }
  \newcommand{\vkk}{\frac{ \vK}{\vKsq} }

  \newcommand{\R}{ \psi }
  \newcommand{\T}{ \Theta }
  \newcommand{\Z}{ \zeta }
  \newcommand{\A}{ \alpha }
  \newcommand{\U}{ u }
  \newcommand{\ve}{ \vect{e} }

  \newcommand{\vlt}{ \vect{e}_\Theta }
  \newcommand{\vlz}{ \vect{e}_\zeta }
  \newcommand{\gr}{ \grad \R }
  \newcommand{\gt}{ \grad \Theta }
  \newcommand{\gz}{ \grad \zeta }
  \newcommand{\ga}{ \grad \alpha }
  \newcommand{\gu}{ \grad \U }
  \newcommand{\dr}[1]{ \frac{\partial #1}{\partial \R} }
  \newcommand{\dT}[1]{\frac{\partial #1}{\partial \Theta}}
  \newcommand{\dz}[1]{\frac{\partial #1}{\partial \zeta}}
  \newcommand{\dU}[1]{\frac{\partial #1}{\partial \U}}
  \newcommand{\drs}[1]{ \frac{\partial^2 #1}{\partial \R^2} }

  \newcommand{\grr}{ g^{\R \R} }
  \newcommand{\grt}{ g^{\R \Theta} }
  \newcommand{\grz}{ g^{\R \zeta} }

  \newcommand{\fr}{ \lp \R \rp}
  
  \newcommand{\frtz}{ \lp \R,\T,\Z \rp}

  \newcommand{\fluxav}[1]{\la #1 \ra}
  \newcommand{\thetaav}[1]{\la #1 \ra_\T}

  \newcommand{\gradt}{\grad_t} 
  \newcommand{\gct}{ \gradt \Theta }	
  \newcommand{\gcz}{ \gradt \zeta }
  \newcommand{\gca}{ \gradt \alpha }
  
  \newcommand{\pb}{\beta} 
  \newcommand{\pbh}{\hat{\beta}}
  \newcommand{\pT}{T} 
  \newcommand{\pZ}{Z} 
  \newcommand{\pg}{g} 
  \newcommand{\ph}{h} 
  \newcommand{\pd}{d} 
   
  \newcommand{\pG}{G} 
   
  \newcommand{\pD}{D} 
  \newcommand{\Fp}{ F^\prime }				

  \newcommand{\sig} {\sigma}
  \newcommand{\sigp}{ \frac{\sig}{P^\prime } }

  \newcommand{\curv}{\vect{\kappa}}
  \newcommand{\curvr}{\kappa_\R}
  \newcommand{\curvc}{\kappa_\C}
  \newcommand{\curvn}{\kappa_n}
  \newcommand{\curvg}{\kappa_g}
  \newcommand{\shear}{\mathcal{S}}
  \newcommand{\residual}{\mathcal{R}}
  \newcommand{\gra}{\frac{g^{\R\A}}{\grr}}

  \newcommand{\ntorsion}{\tau_n}
  \newcommand{\normal}{\hat{\vect{n}}}
  \newcommand{\bhat}{\hat{\vect{b}}}
  \newcommand{\binormal}{\normal \times \bhat}

  \newcommand{\eps}{\epsilon}
  \newcommand{\order}[1]{{\mathcal{O}\lp \eps^{#1} \rp}}
  \newcommand{\pone}{{\lp1\rp}}
  \newcommand{\ptwo}{{\lp2\rp}}
  \newcommand{\pthree}{{\lp3\rp}}
  
  \newcommand{\vb}{\vect{b}}
  \newcommand{\vxi}{\vect{\xi}}
  
  \newcommand{\br}{b^\R}
  \newcommand{\bb}{b^B}
  \newcommand{\bc}{b^\C}
  \newcommand{\bj}{b^J}
  \newcommand{\bk}{b^K}
  \newcommand{\xr}{\xi^\R}
  \newcommand{\xb}{\xi^B}
  \newcommand{\xc}{\xi^\C}
  \newcommand{\xj}{\xi^J}
  \newcommand{\xk}{\xi^K}
  \newcommand{\abr}{\la \br \ra}
  \newcommand{\abb}{\la \bb \ra}
  \newcommand{\abc}{\la \bc \ra}
  \newcommand{\abj}{\la \bj \ra}
  \newcommand{\abk}{\la \bk \ra}
  \newcommand{\axr}{\la \xr \ra}
  \newcommand{\axb}{\la \xb \ra}
  
  \newcommand{\axj}{\la \xj \ra}
  \newcommand{\axk}{\la \xk \ra}
  
  \newcommand{\Sv}{ {\Psi} }
  \newcommand{\Xv}{ {\Xi} }
  \newcommand{\Bv}{ {\Upsilon} }
  \newcommand{\Gv}{ {\Gamma} }

  \newcommand{\eog}{ \frac{\eta}{\gamma}}

  \newcommand{\absq}{\la B^2 \ra}
  \newcommand{\acsq}{\la \C^2 \ra}
  \newcommand{\asbsq}{\la \sig B^2 \ra}
  \newcommand{\ascsq}{\la \sig \C^2 \ra}
  \newcommand{\asqcsq}{\la \sig^2 \C^2 \ra}
  \newcommand{\aoobsq}{\la \frac{ 1}{B^2 } \ra}
  \newcommand{\M}{ {M} }
  
  \newcommand{\X}{ \bar{x} }
  \newcommand{\EC} {\vC \cdot \curl{\vK}}

  \newcommand{\CE}{ \frac{-\acsq \Pp}{\Lambda^2} \lb \fluxav{\EC} 
				+ \frac{\Lambda}{\Pp} \frac{\asbsq}{\absq} \rb }
  \newcommand{\CF}{ \frac{\acsq}{\Lambda^2} \lb P^{\prime 2} \aoobsq 
  	+ \asqcsq - \frac{\ascsq^2}{\acsq} \rb }
  \newcommand{\CH}{ \frac{\acsq}{\Lambda} \lb \frac{\asbsq}{\absq} - \frac{\ascsq}{\acsq} \rb }
  \newcommand{\CM}{\frac{\acsq}{P^{\prime 2}} \lb P^{\prime 2} \la \frac{1}{C^2} \ra 
			   + \la \sig^2 B^2 \ra - \frac{\asbsq^2}{\absq} \rb }
  \newcommand{\CG}{ \frac{\absq}{\specheat P \M}  }
  \newcommand{\CK}{ \frac{\Lambda^2 R}{P^{\prime 2} \M}  }

\begin{document}

\title{
  The Relationship between Flux Coordinates and Equilibrium-based Frames of
  Reference in Fusion Theory
}


\author{S.E. Kruger}
\affiliation{Tech-X Corporation}
\email{kruger@txcorp.com}
\author{J.M. Greene}
\affiliation{
      Deceased. The original idea of the $JK$ reference frame and the
      development of the new annihilation operator is due strictly to Dr. J.M. Greene.}

\begin{abstract}

The properties of two local reference frames based on the magnetic field and the
current density are investigated for magnetized plasmas in toroidal geometry
with symmetric angle.  The magnetic field-based local frame of reference has
been well-studied for example by Dewar and colleagues [Phys. Fluids 27, 1723
(1984)] An analogous frame based on the current density vector is possible
because it is also divergence free and perpendicular to the gradient of the
poloidal flux.  The concept of straightness of a vector is introduced and used
to elucidate the Boozer and Hamada coordinate systems.  The relationship of
these local frames to the more well-known Frenet frame of reference, which
specifies a curve in terms of curvature and torsion, is given.  As an example of
the usefulness of the these formal relationships, we briefly review the ideal
MHD theory and their use.  We also present a new annihilation operator, useful
for eliminating shorter time scales than the time scale of interest, for
deriving the inner layer equations of Glasser, Greene, and Johnson [Phys. Fluids
18, 875 (1975)].  Compared to the original derivation that is based on the local
frame of reference in terms of the magnetic field, the new annihilation operator
that is based on the local frame of reference in terms of the current density
simplifies the derivation.

\end{abstract}

\pacs{52.30.-q, 52.65.-y, 52.35.-g, 52.55.Tn, 52.40.Hf, 02.60.Lj, 52.35.Vd}


\preprint{DRAFT}	

\date{\today}

\maketitle
\section{Introduction}

High density, magnetized plasmas, such as tokamak or stellarator plasmas, are
characterized by a large  number of temporal and spatial scales.  Laboratory
magnetized plasmas have had great success in creating plasmas with confinement
times much longer than the shortest time scales.  Understanding plasma behavior
on these longer times requires analytic, and computational, techniques for
stepping over the shortest time scales to study the time scales of interest.  A
key issue for magnetized plasmas is resolving the anisotropy of the magnetic
field.  For example, examination of the Braginskii transport equations
reveals different transport time scales for the directions parallel to the
magnetic field, perpendicular to the magnetic flux surfaces, and the binormal
direction that is perpendicular to both directions~\cite{Braginskii}.  This
anisotropy arises consistently in fluid and kinetic theories.

A common method for analytically handling this anisotropy is to use a local
frame of reference.  By frame of reference, we mean the development of a local
coordinate system at a point in space.  For example, the three directions above
enable the development of a set of basis vectors for a local coordinate system.
This local frame of reference has long been used in fluid
theory~\cite{GreeneJohnson62,CGJ} in combination with the use of a flux
coordinates~\cite{GreeneJohnson62} appropriate for toroidal systems.
Dewar and colleagues~\cite{Dewar84}, hereafter referred to as DMS,
greatly improved the understanding of this local reference frame by formalizing
many of the relationships between flux coordinates and the local frame that are
commonly used in analytic derivation.  This frame of reference is also implicit
in much of the work of Stix~\cite{Stix} with the ``Stix
frame''~\cite{wright_thesis1998} being a common tool in RF theory.  That frame,
like the more well-known Frenet frame for arbitrary parameterized curves, is
expressed in terms of dimensionless vectors.  As we shall see, for most of the
theory calculations, using dimensional local basis vectors is preferred.

Flux coordinate systems make use of two properties of the equilibrium magnetic
field.  The first is that the magnetic field is divergence-free.  The second is
the equilibrium relationship: 
\begin{equation}
   \vJ \times \vB = \grad P.
   \label{magEquil}
\end{equation}
which is the zeroth-order force balance in a magnetized
plasma~\cite{HazeltineMeiss}. 
Dotting both sides with $\vB$ shows that $\vB \cdot \grad P = 0$.
This combined with symmetry allows the
definition of a flux function such that $\vB \cdot \grad \psi =0$.
Quasineutrality enables the charge continuity equation to be written as
$\dive{\vJ}=0$.  The equilibrium relationship also shows that $\vJ \cdot \grad P
=0$ or that $\vJ \cdot \grad \psi = 0$ as well.  Thus, it is obvious that
another local frame of reference, completely analogous to the magnetic
field-based frame is possible using the current density vector.  To our
knowledge, this has never been explored before.  In this paper, we formalize the
relationships between the two reference frames and flux coordinate systems
similar in spirit to the work of the DMS paper.  We also relate this
frame of reference to the Frenet frame from differential geometry of curves,
which expresses a curve in terms of curvature and torsion.   Fusion theory has
favored magnetic shear over torsion, and these relationships are reviewed.

As an example of the usefulness of formalizing the relationships between flux
coordinates and local frames of reference, their usefulness in ideal MHD and
resistive MHD instability theory is reviewed.  Resistive instabilities, which
have a time scale that is a hybrid of the Alfv\`en and resistive diffusion time
scales, is much slower than the Alfv\`en time scales.  Analytically, this requires
going to second order in the ordering, which involves significant algebra,
especially in studying instabilities in toroidal geometry with complicated flux
surface shapes.  The rest of the paper reviews the derivation of the inner layer
equations in toroidal geometry.  The first discussion of how to derive these
equations is contained in the appendix of Johnson and
Greene~\cite{JohnsonGreene67} in 1967, but they were not fully derived and
published until Glasser, Greene, and Johnson (GGJ) in 1967.  Unfortunately, that
paper only focused on the analysis of these equations and only presented them.
Part of the point of this paper is that the derivation of these equations
contains pedagogical value, and a new method of deriving these equations is
presented.

The rest of the paper proceeds as follows.  First, we introduce the concept of
``equilibrium-based frame of references'' and introduce our new frame.  Next,
following the work of DMS, we provide the mapping of
the equilibrium-based frames of references to the flux coordinate systems.  We
then briefly discuss the relationship of equilibrium reference frames to the
Frenet frame and common geometric quantities.   With the mathematical discussion
complete, we then discuss the ideal MHD formulas, and derivation of the inner
layer equations in toroidal geometry.  Key to the derivation is the use of
annihilation operators~\cite{CGJ,GGJ}.  For the derivation of the inner layer
equations in toroidal geometry, we find that a new annihilation operator, based
on the new equilibrium-based coordinate system, is more useful in toroidal
geometry.  Finally, we conclude by discussing the implications of the
derivation, and the relationship to computational approaches.
    \newpage \clearpage
\section{Equilibrium-Based Frames of Reference}

Frames of references are local coordinate systems that vary throughout space.
An example would be the rotating frame of reference describing the forces felt
by a person on a merry-go-round commonly used to explain the difference between
centripetal and centrifugal forces.  Another well-known example from
mathematical physics is the Frenet frame of reference which constructs a local
set of basis vectors along a curve in terms of the tangent of the curve, the
curvature vector which is normal to the tangent, and the binormal vector created
from the cross-product of both the tangent and curvature.  This creates a local
orthonormal set of basis vectors at a given point, but it is not a general
coordinate system because it cannot be used to label an arbitrary point in
space.

As discussed in the introduction, the most common frame of reference has the
magnetic field, the gradient of a flux function, and the binormal direction as
the local basis vectors.  We term this frame the $BC$ frame and define it as:
\begin{equation}
	 \vR,\ \  \vB,\ \  \vC 
	\label{BCcontra}
\end{equation}
where $\psi$ is the poloidal magnetic flux divided by $2 \pi$, and
$\vC$ is constructed to make the coordinate system orthogonal:
\begin{equation}
	  \vC \equiv \frac{\vR \times \vB}{\vRsq}
  \label{CDef}	
\end{equation}
Unlike the Stix frame which normalizes the basis vectors so that they are
unit-less (and orders the parallel direction last for
convenience)~\cite{wright_thesis1998}, the normalizations chosen here seem to be
the most useful as will be shown.

In this paper, we define a new equilibrium-based frame of reference, the
$JK$ frame
\begin{equation}
	 \vR,\ \   \vJ,\ \   \vK 
	\label{JKcontra}
\end{equation}
where $\vK$ is likewise defined as
\begin{equation}
	  \vK \equiv \frac{\vR \times \vJ}{\norm \vRsq}.
  \label{KDef}	
\end{equation}
The $\Pp$ denotes the derivative of the pressure with $\psi$.  Our use of this
normalization is to make $\vK \cdot \vB = 1$ which is useful.
Although using the other vector in the equilibrium relation, $\vJ$, as the basis
of an equilibrium-based frame seems obvious, to our
knowledge it has never been studied.  In this paper, we show how for
long wavelength instabilities at least, it is useful to explicitly treat
it as such.

Using vector identities, it is easily seen that the other vector
components within each system may be expressed by
\begin{align}
									&&
	  \vB = \vC \times \vR 
	  ,\ \ \ \ \ \ \ \ 
	  \vR = \frac{\vB \times \vC}{\vCsq}
  \label{BCcoord}	
									\\  &&
	  \vJ = \norm \vK \times \vR 
	  ,\ \ \ \ \ \ \ \ 
	  \vR = \frac{\vJ \times \vK}{\norm \vKsq}
  \label{JKcoord}	
\end{align}

To express an arbitrary vector in terms of these local basis vectors, we borrow
notation from the curvilinear flux coordinate systems for simplicity.  If one
considers these two vector systems as ``contravariant-like'' basis vectors,
$\ve^i, i \in 1,2,3$, then the concomitant ``covariant-like'' basis vectors,
$\ve_i$ are given by 
$\ve_1 = \epsilon^{123} \jac \ve^2 \times \ve^3$
where 
$\jac = (\ve^1 \cdot \ve^2 \times \ve^3)^{-1}$ 
is the Jacobian of the
system, and $\epsilon^{123}$ is the Levi-Cevita symbol.  
The Jacobian is used as a normalization so
that $\ve^i \cdot \ve_j = \delta^i_j$ where $\delta^i_j$ is the Kronecker
delta.  Any vector is represented in terms of these basis vectors by the
following representation 
$\vA = \sum_i A^i \ve_i = \sum_i A_i \ve^i,\ i \in \{1,2,3 \}$.
The ${A^i}$'s are the contravariant components and the $A_i$'s are 
the covariant components.  
By the above relations, the components may clearly be seen to be given by 
$A^i = \vA \cdot \ve^i$ and $A_i = \vA \cdot \ve_i$.  

Using this notation for the basis vectors above,
the inverse Jacobians of the two systems are
\begin{align}
									&&
	  B^2 = \vCsq \vRsq = \vR \cdot \vB \times \vC 
									\\  &&
	  \frac{J^2}{\norm} = \vKsq \vRsq = \vR \cdot \vJ \times \vK 
\end{align}
so that the covariant-like basis vectors are:
\begin{align}
	 \vrr,\ \ \ \  \  \vbb , \ \ \ \   \vcc &&
									\\ 
	\vrr,\ \ \ \  \vjj , \ \ \ \   \vkk. &&
\end{align}
The use of $\vRsq$ in the definition of the binormal vectors $\vC$ and $\vK$ is
to make these covariant-like basis vectors have this consistent form.  An
arbitrary vector $\vect{f}$ then can be written as
\begin{eqaligned}
	  \vect{f} &= f^\psi \vrr + f^B \vbb + f^C \vcc
									\\  
	          &= f^\psi \vrr + f^J \vjj + f^K \vkk.
\end{eqaligned}
Both the Stix frame and the Frenet frame use basis unit vectors and decompose
vectors into components that have the same units.  Using these
``contravariant-like'' components is perhaps un-intuitive, but they have long been
used in stability analyses~\cite{CGJ}.  As we show in
\refsec{InnerLayerEquations}, this is because they identify the correct
components that equilibrate quickly due to compressional Alfv\`en waves.

Using force balance $\vJ \times \vB = \grad P$, and defining a variable
$\sigma$ for the parallel current, the relationships between the 
$BC$ and $JK$ frames are:
\begin{align}
	  \vJ = \sig \vB - \norm \vcc,
	  &\ \ \ \ &
	  \vK = \sigp \vC + \vbb
	\label{JKBC}
									\\  
	  \vB = \sig \vBsq \vjj + \vkk, 
	  &\ \ \ \ &
	  \vC = \frac{\sig \vCsq}{\norm} \vkk - \norm \vjj
	\label{BCJK}
\end{align}
The metrics of this transformation are
\begin{eqaligned}
	  \vJ \cdot \vB = \sig B^2, 
	  &\ \ \ \ \ \ &
	  \vJ \cdot \vC = -\norm
									\\  
	  \vK \cdot \vB = 1, 
	  &\ \ \ \ \ \ &
	  \vK \cdot \vC = \frac{\sig \vCsq}{\Pp}.
\end{eqaligned}
The angle between the coordinate systems then is related to $\sig/P'$, which is
a flux function in cylindrical geometry but not in toroidal geometry.
To restore symmetry to the coordinate systems, one
could also use $\Pp$ in the definition of $\vC$, or equivalently, use
$\grad P$ as the radial vector, which has a certain elegance since all
three vectors of the equilibrium relation would then be used
as the basis vectors.
The choice of normalizations used here seem to be the easiest to
use in practice however.
The contravariant vector components are related by
\begin{align}
	  f^\C = \frac{\norm}{J^2} (\sig  B^2 f^K - \norm f^J),
	  &\ \ \ \ &
	  f^B =  \frac{1}{K^2} (\sig \C^2 f^J + f^K)
\label{BCcomps}
									\\  
	  f^K = \frac{1}{\norm B^2} (\sig  B^2 f^\C + \norm f^B),
	  &\ \ \ \ &
	  f^J =  \frac{1}{\C^2} (\sig \C^2 f^B - \norm f^\C).
\label{JKcomps}
\end{align}
The metrics presented here appear in the Mercier criterion; thus, one
sees even here that stability can be viewed as a complicated function of
the angle between these two frames.

We summarize the expressions in this section in Table~\ref{tab:coordinateSummary}.
While these frames of reference are useful, it is necessary to make
use of geometry-based coordinate systems to study the physics in the
appropriate magnetic configuration.  The rest of this paper focuses on
the doubly-periodic systems relevant to tokamaks and symmetric
stellarators.

    \newpage \clearpage
\section{Flux Coordinates and the Relationship to Equilibrium-based Frames}

While the local frames of reference are useful for getting at the physics of the
problem, they do not take advantage of the periodicity of the toroidal systems
that we wish to study.  Following the work of DMS, the relationship between the
equilibrium-based frames of reference and the geometry-based coordinate systems
is explicitly examined.  We start by briefly reviewing flux
coordinates~\cite{HazeltineMeiss}, labeled in our notations as the $\R, \T, \Z$
system.  In this paper, only systems with at least one degree of symmetry are
considered with $\Z$ being the ignorable coordinate.  This includes helically
symmetric systems.

We assume nested flux surfaces and will consider various averages
related to those surfaces.
 A {\em volume average} of a quantity $Q$ at a surface $\psi$ is denoted as
 \begin{equation}
 	\la Q \ra_V \defn \frac{1}{V(\R)} \int Q\ dV
 	            = \frac{1}{V(\R)} \int_0^{2 \pi}  \int_0^{2 \pi}  
 			   \int_{\R_{min}}^{\R} Q\ \jac d\R' d\T d\Z
 \end{equation}
 where $V(\R)$ is the volume enclosed by a flux surface,
 $V = \int \int \int \jac d\R' d\T d\Z$.
 The {\em flux-surface average} of a quantity $Q$ 
 at a surface $\psi$ is denoted as
 \begin{equation}
 	\fluxav{Q} = \frac{1}{V'(\R)} \int_0^{2 \pi}  \int_0^{2 \pi}  
 			   Q\ \jac d\T d\Z
   \label{FSA}
 \end{equation}
 where $V'(\R)$ is the derivative of $V$ with respect to $\R$,
 $V' = \oint  \oint  \jac d\T d\Z$.
 Also, we define the {\em theta average} of a quantity $Q$ as
 \begin{equation}
 	\thetaav{Q} = \frac{1}{2 \pi } \int_0^{2 \pi}  Q\ d\T d\Z.
 \end{equation}
 For axisymmetric quantities, this average is closely related to the
 flux-surface average; i.e., if $Q$ is independent of $\Z$, then 
 $\thetaav{\jac Q} = \Vp/(4 \pi^2) \fluxav{Q}$.

\subsection{Properties of divergence-free fields perpendicular to flux surfaces}
Flux coordinates are usually derived using the properties of the 
divergence-free nature of the magnetic field along with the definition of flux
surface as a surface perpendicular to the magnetic field.  Our equilibrium-based
local frames of reference include two other divergence-free vectors, $\vJ$ and
$\vRsq vC$, so it is useful to first understand general properties of
vector-free fields perpendicular to flux surfaces.  
The divergence-free property of $\vRsq vC$ can se seen using $\vJ \cdot \vR = 0$:
\begin{equation}
	  \vR \cdot \vJ = \vR \cdot \curl{\vB} = \dive{\vB \times \vR}
	                = \dive{\vRsq \vC} = 0.
\label{CIdentity}
\end{equation}
No similar relation can be found for $\vK$.  This makes this vector
fundamentally different from the other basis vectors.

Consider the general properties of a vector $\vV$ with
$\dive{\vV}=0$ and $\vV \cdot \vR = 0$.  These properties allow
$\vV$ to be written in a Clebsch representation:
\begin{equation}
		\vV = \grad \pb_V \times \vR.
\end{equation}
The most general form for $\pb_V$ in toroidal geometry is~\cite{Boozer81}:
\begin{equation}
		\pb_V = \pT_V\fr \Z - \pZ_V\fr \T + \pbh_V\frtz
   \label{pbv_eq}
\end{equation}
where $\pbh_V$ is a periodic function of $\T$ and $\Z$.  This 
allows $\vV$ to be written as
\begin{eqaligned}
		\vV &= \pT_V \gz \times \gr 
			+ \pZ_V \gt \times \gr 
			+ \grad \pbh_V \times \vR
									\nonumber \\  
		    &= \jaci \lb \lp \pT_V + \dz{\pbh_V}\rp \vlt 
		                - \lp \pZ_V - \dT{\pbh_V}\rp \vlz \rb
	\label{contraV}
\end{eqaligned}
which is the covariant form for $\vV$.  

Using the expressions for the contravariant components of $\vV$ in 
Eq.\ (\ref{contraV}), the poloidal flux of $\vV$ may be easily calculated:
\begin{eqaligned}
	\Gamma_V^\T\fr
            &\defn \frac{1}{(2 \pi)^2} 
				     \int \oint \vV \cdot \gt\ \jac d\Z d\R
									\nonumber \\  
		    &=  \frac{1}{2 \pi} \int \pT_V d\R +
			   \frac{1}{(2 \pi)^2} \int \oint \dz{\pbh_V}d\Z d\R.
\end{eqaligned}
The last term is zero due to periodicity.  Following a similar procedure
for the toroidal flux, 
$\Gamma_V^\Z\fr \defn 1/(2 \pi)^2 \int \oint \vV \cdot \gz\ \jac d\T d\R$,
one obtains the relationships
\begin{equation}
	\pT_V = 2 \pi {\Gamma_V^\T}^\prime;\ \ \ \ \ \ \pZ_V = 2 \pi {\Gamma_V^\Z}^\prime.
	\label{FluxRelations}
\end{equation}
which relates the contravariant components to the fluxes. 

The fact that $\vV \cdot \vR = V^\R = 0$ allows the concept of
{\em straightness} to be defined.  A vector $\vV$ with zero
$\R$ contravariant component is said to be
{\em straight} if for each flux surface defined by $\R$, the ratio of the
remaining contravariant components is constant; i.e., $\vV$ is straight
if $V^\Z/V^\T = f\fr$.
Looking at the contravariant components for $\vV$ in
Eq.\ (\ref{contraV}),  it is apparent that
this is true, if and only if $\pbh_V=0$.  In this case,
\begin{equation}
	\frac{V^\Z}{V^\T} = 
	\frac{\pZ_V}{\pT_V} = 
	\frac{{\Gamma_V^\Z}^\prime}{{\Gamma_V^\T}^\prime} = 
	f\fr.
\end{equation}
Calculations involving a straight vector are obviously greatly simplified
because of the simplification of the contravariant components to flux functions.

To convert between $\vB$ and $\vC$, and $\vJ$ and $\vK$, the
following identity for the concomitant vector,
$\vR \times \vV/\vRsq$,  is useful:
\begin{eqaligned}
	\frac{\vR\times\vV}{\vRsq} &= \vR \times \lp \grad \pb_V \times \vR \rp
									\nonumber \\  
	                           &= \grad \pb_V 
					   	 - \frac{\grad \pb_V \cdot \vR}{\vRsq} \vR
									\nonumber \\  
	                           &\defn \gradt \pb_V 
\end{eqaligned}
where in the last line a new operator is introduced for convenience.
The $\gradt$ operator with respect to $\vR$ is analogous to the more familiar
$\grad_\perp$ operator \cite{Kruger99} with respect to $\vB$.  The $t$ notation
is used because $\gradt f$ is tangential to the constant $\R$ surfaces.
Using this notation and the equation for $\pb_V$, Eq.\ (\ref{pbv_eq}), the
concomitant vector may be written as
\begin{equation}
	\frac{\vR\times\vV}{\vRsq} = \lp \pT_V + \dz{\pbh_V} \rp \gcz 
					   - \lp \pZ_V - \dT{\pbh_V} \rp \gct.
	\label{ConcomVec}
\end{equation}
Since $\vV$ has a convenient contravariant form, the concomitant vector
will have a convenient covariant form, especially if $\vV$ is straight.

After the development of this formalism, the calculations for the
specific cases of $\vV \in \left\{ \vB,\vJ,\vRsq \vC \right\}$ are greatly simplified.
In particular, our goal is to find expressions for 
$\left\{ \pT,\pZ,\pbh \right\}_{\left\{B,C,J \right\} }.$

\subsection{Flux coordinates}
In this paper, only {\em flux coordinate systems} where $\vB$ is straight
($\pbh_B=0$) are considered.  Because the radial coordinate has been
chosen to be $\psi \defn 2 \pi \Gamma_B^\T$, the components of $\vB$ have
the nice form
\begin{equation}
			\pT_B=1;\ \ \ \ \ \ \pZ_B=q\fr
\end{equation}
where 
\begin{equation}
	q\fr \defn \frac{B^\Z}{B^\T} = 
		\frac{{\Gamma_B^\Z}^\prime}{{\Gamma_B^\T}^\prime}.
\end{equation}
is the safety factor.  This simple form for $\vB$ allows $\vC$ to be
written in the contravariant form 
\begin{equation}
   \vC= \gcz - q \gct
   \label{C_contra}
\end{equation}
using Eq.\ (\ref{ConcomVec}).

Considering the covariant form of $\vRsq \vC$ from Eq.\ (\ref{contraV})
and using Eq.\ (\ref{ConcomVec})
to find the contravariant form of $\vB$ (Eq.\ (\ref{BCcoord})) yields
\begin{equation}
	\vB = -\lp \pT_C + \dz{\pbh_C} \rp \gcz 
					   + \lp \pZ_C - \dT{\pbh_C} \rp \gct.
	\label{CoVarB}
\end{equation}
At this point, the ignorability of $\Z$ is used to simplify
the $\Z$ covariant component of $\vB$. We define
\begin{equation}
	F\fr \defn -\pT_C = -\jac C^\T = -2 \pi {\Gamma_C^\T}^\prime
\end{equation}
In axisymmetric systems, $F$ is known as the toroidal flux function
$F=R B_{toroidal}$.  To find a useful expression for $\pZ_C$, the covariant
form of $\vB$ (Eq.\ (\ref{CoVarB})) is dotted with the contravariant 
form of $\vB$ to give
\begin{equation}
	B^2 = \jaci \lp \pZ_C - \dT{\pbh_C} \rp + \jaci qF.
\end{equation}
For convenience, we define a new variable, $\pg$,
\begin{equation}
	\pg \defn \pZ_C - \dT{\pbh_C}  = \jac B^2 - qF.
	\label{gdef}
\end{equation}
Using the $\T$ average allows us to annihilate the second term and solve:
\begin{eqaligned}
	\pZ_C = \thetaav{\pg} \defn \pG\fr &= \thetaav{\jac B^2} - qF 
	                    =  \Vpo \fluxav{B^2}  - qF;
									\nonumber \\  
	\dT{\pbh_C} &= \thetaav{\jac B^2} - \jac B^2
	        = \Vpo \fluxav{B^2} - \jac B^2.
\end{eqaligned}
Evidently, to make $\vC$ straight, $\jac B^2$ needs to be a flux function.
A coordinate system with this property is known as the 
{\em Boozer flux coordinate system} \cite{Boozer81,Boozer82}.
Using these definitions, the covariant form of $\vB$ can be written as
$\vB = F \gcz + \pg \gct$.

Taking the curl of $\vB$ to give the current density gives
\begin{equation}
	\vJ = -\Fp \gz \times \gr 
	      - \lb \dr{\pg} 
		      - \dT{}\lp F \frac{\grz}{\grr} +\pg \frac{\grt}{\grr} \rp
		  \rb \gt \times \gr.
     \label{Jform1}
\end{equation}
The expression for $\pT_J$ is simply $\pT_J= -\Fp$; thus, the
function $F$ introduced above is proportional to the poloidal
current enclosed by a flux surface as seen from 
Eq.\ (\ref{FluxRelations}) ($F=-2 \pi \Gamma_J^\T$).  Similarly, one can
show $G\fr = \thetaav{g} =2 \pi \Gamma_J^\Z$ is proportional to the
toroidal current enclosed by a flux surface.

The above expression for $\vJ$ was derived using Ampere's law.
The expression for $\pZ_J$
can be simplified by deriving another covariant form of $\vJ$ using 
Eq.\ (\ref{JKBC}) and the covariant forms for $\vB$ and $\vC$ already derived.  
Using the covariant forms yields
\begin{equation}
	\vJ =  \lp   \sigma + \frac{\Pp F}{B^2}   \rp \gz \times \gr 
		-\lp q \sigma - \frac{\Pp \pg}{B^2} \rp \gt \times \gr.
     \label{Jform2}
\end{equation}
Equating the $\T$ contravariant components yields an expression for the
parallel current:
\begin{equation}
		\sigma = -\Fp - \frac{\Pp F}{B^2}. 
\end{equation}
This relation allows us to rewrite the $\Z$ contravariant component of $\vJ$ in
Eq.\ (\ref{Jform2}) as
\begin{equation}
	      \ph \defn \pZ_J - \dT{\pbh_J} = 
			q \sigma - \frac{\Pp \pg}{B^2} =
			-\lp \Pp \jac + q \Fp \rp
	\label{hdef}
\end{equation}
where $\ph$ is defined analogously to $\pg$.
Using the $\T$ average again allows us to annihilate the second term and solve
for each term:
\begin{eqaligned}
	\pZ_J = \thetaav{\ph} = \pG^\prime\fr 
	                      &= -\lp \Pp \thetaav{\jac} + q \Fp \rp 
	                       =  -\lp \Vpo \Pp           + q \Fp \rp 
				     = 2 \pi \Gamma_J^\Z;
									\nonumber \\  
	\dT{\pbh_J} &= \Pp \lp \jac - \thetaav{\jac} \rp
	             =  \Pp \lp \jac - \Vpo \rp.
\end{eqaligned}
Evidently, to make $\vJ$ straight, the Jacobian needs to be a flux function.
A coordinate system with this property is known as the 
{\em Hamada flux coordinate system}.

Equating the $\Z$ contravariant components yields 
\begin{equation}
		\dr{\pg} + \dT{}\lp F \frac{\grz}{\grr} +\pg \frac{\grt}{\grr} \rp
		=  \ph
	\label{GenGS}
\end{equation}
which is a generalized Grad-Shafranov equation.

The contravariant form of $\vJ$ can be derived using  
Eq.\ (\ref{JKBC}) and the covariant forms for $\vB$ and $\vC$ already derived.  
Using the contravariant forms yields
\begin{equation}
	\vJ =  \lp \sigma   F - \frac{\Pp}{C^2}   \rp \gcz
		+\lp \sigma \pg + \frac{q \Pp}{C^2} \rp \gct.
     \label{Jform3}
\end{equation}
Using the contravariant and covariant forms of $\vJ$, expressions for $\vK$ 
may be easily derived using Eq.\ (\ref{KDef}).  The results are
\begin{equation}
    \vK = \frac{-1}{\Pp \vRsq} \lb \lp \sigma   F - \frac{  \Pp}{C^2} \rp \gz \times \gr 
	     \lp \sigma \pg + q \frac{\Pp}{C^2} \rp \gt \times \gr \rb
  \label{K_co}
\end{equation}
and 
\begin{equation}
	 \vK = \frac{-1}{\Pp} \lb \Fp \gcz  + \ph  \gct \rb
  \label{K_contra}
\end{equation}

We summarize the expressions which were derived above for all four of our basis
vectors in Table~\ref{tab:coordinateSummary}.

\subsection{Helical angle coordinates}
In addition to the $\R, \T, \Z$ coordinate system, 
another useful coordinate system is the $\R, \T, \A$ coordinate system 
where $\A$ is the Clebsch angle for $\vB$
\begin{equation}
		\A \defn \pb_B = \Z - q \T.
	\label{ADef}
\end{equation}
such that $\vB = \ga \times \gr$.  
If $\Z$ is an ignorable coordinate, then $\A$ is an
ignorable coordinate also.  
The primary advantage of
this coordinate system is that the tangential gradient of
$\A$ is equal to $\vC$ as can easily be seen by 
Eq.\ (\ref{C_contra}):
\begin{equation}
		\vC = \gca = \ga - \gra \vR.
	\label{CeqA}
\end{equation}
Expressions for the basis vectors using this coordinate system
can be easily evaluated.  
We summarize the expressions which were derived above for all four of our basis
vectors in Table~\ref{tab:coordinateSummary}.

Examination of the equation for $\gca$, Eq. (\ref{CeqA}), in more detail, 
shows that the last term may be written as
\begin{eqaligned}
		\gra &= \frac{1}{\vRsq} \lp \gz \cdot \vR 
							- \grad(q \T) \cdot \vR \rp
				= -\lp q'\T
				    +\frac{q g^{\RR\T}-g^{\RR\Z}}{g^{\RR\RR}} \rp
									\nonumber 	\\
				&= -\lp q'\T + \residual(\R,\T) \rp.
	\label{graEq}
\end{eqaligned}
The $\residual$ term is known as the residual shear and will be discussed in the
next section.
As evidenced by the $q'\T$ term in the previous equation, $\A$ is not
a periodic coordinate.  While this coordinate system is still useful
for short wavelength modes, it is often more useful to have a coordinate
system which is periodic, yet closely related to $\A$.  We define
the coordinate $\U$ as
\begin{equation}
		\U \equiv \Z - q_s \T = \A + (q - q_s) \T
	\label{UDef}
\end{equation}
where $q_s$ is the value of $q$ at a particular surface.  
For a vector $\vF \perp \vR$, 
$\vF \cdot \gu = \vF \cdot \ga + (q - q_0) \vF \cdot \gt$ which
may be used to find the $\U$  contravariant components and following a
similar derivation as above, the covariant components as well.
The summary of these relationships is shown in vectors in
Table~\ref{tab:coordinateSummary} as well.
\subsection{Special cases of flux coordinate systems}
Until this point, $\T$ and $\Z$ are only specified such that the 
magnetic field lines are straight: 
\begin{equation}
	\frac{\vB \cdot \gz}{\vB \cdot \gt} = q(\R).
\label{StraightB2}
\end{equation}
This is one constraint on two quantities.  
To completely specify a straight-field line coordinate system, another 
requirement is needed; generally the Jacobian is specified.  

Previously we mentioned two specific coordinates, the Boozer and Hamada
coordinate systems, and here we describe their Jacobians and properties.  Recall
that Hamada coordinates specify the constraint that $\vJ$ is straight,
\begin{equation}
	\frac{\vJ \cdot \gz}{\vJ \cdot \gt} = f(\R),
\label{StraightJ}
\end{equation}
and Boozer coordinates specify that the magnetic binormal vector is
straight~\footnote{A useful mnemonic:  Allen Boozer made C straight as simple
as ABC.  Hamada Is J straight as simple as HIJ.},
\begin{equation}
	\frac{\vC \cdot \gz}{\vC \cdot \gt} = f(\R).
\label{StraightC}
\end{equation}
In addition to these two coordinate systems, we also discuss the ``symmetry
coordinates''.  Practical aspects related to calculating the coordinate systems
numerically may be found in DMS~\cite{Dewar84}.

\subsubsection{Hamada Coordinate System}
Hamada coordinates are elegant because it makes the current density
field lines straight as well, thus continuing the analogous relationships
between $\vB$ and $\vJ$.  From Eq.\ (\ref{hdef}),  we see that 
\begin{equation}
  \frac{\vJ \cdot \gz}{\vJ \cdot \gt} = \frac{h}{F'} = q + \jac \frac{\Pp}{F'}
\label{JztRatio}
\end{equation}
To make these straight, we need to make the Jacobian a flux-function.
This flux function may be found from the relationship:
\begin{equation}
	  \fluxav{\jaci} = \frac{4 \pi^2}{V'}.
\end{equation}
which implies that the Jacobian for the Hamada coordinate system may be written
as
\begin{equation}
	  \jac_H = \frac{V'}{4 \pi^2}.
\end{equation}
\begin{equation}
	\vJ = -F'(\R) \gz \times \grad \R - G'(\R) \gt \times \grad \R.
\end{equation}
with $G'/F'=q + 4 \pi^2/(F' V')$.  
\begin{equation}
	- q F'- G' =  \frac{V'}{4 \pi^2} P'.
\end{equation}
which is the Grad-Shafranov equation in Hamada coordinates (using 
$\R$ as the dependent variable). This relationship  is used extensively in Greene
and Johnson~\cite{GreeneJohnson62}, and in the original derivation of the inner
layer equations of GGJ.

\subsubsection{Boozer Coordinate System}
Boozer coordinates have nice properties because $\vC$ is straight as 
well as $\vB$.
From Eq.\ (\ref{gdef}),  we see that 
\begin{equation}
      \frac{\vC \cdot \gz}{\vC \cdot \gt} = \frac{-g}{F} = q - \frac{\jac B^2}{F}
\label{CztRatio}
\end{equation}
To make these straight, we need to have
\begin{equation}
	  \jac_B B^2 = G(\R)
\end{equation}
To find this flux function, we have
\begin{equation}
	  \fluxav{\frac{\jaci_B}{B^2} B^2} = 
	  \frac{\jaci_B}{B^2} \fluxav{ B^2} = 
	  \frac{4 \pi^2}{V'}
\end{equation}
or
\begin{equation}
	  \jac_B B^2 = G(\R) = \frac{V' \fluxav{ B^2}}{4 \pi^2}.
\end{equation}

\subsubsection{Symmetry Coordinates}

In an axisymmetric system, the Boozer and Hamada surfaces of constant $\Z$ will
not correspond to surfaces of constant $\phi$ where $\phi$ is the ``symmetry''
angle.  That is, given a cylindrical $R, \phi, Z$ coordinate system, one is
interested in the case of $\Z = -\phi$~\footnote{The minus sign is to enable the
      right handed-ness to remain in the $(R,\phi,Z)$ and $(\R,\T,\Z)$
coordinate systems.}

In this coordinate system, $\gz \cdot \gz = 1/R^2$, $\gz \cdot \gr = 0$, and
$\gz \cdot \gt = 0$.  This means that the covariant and contravariant components
are parallel:
\begin{equation}
      \gt \times \gr = -\jaci \vlz = \jaci R^2 \gz
\end{equation}
which allows us to write the magnetic field as
\begin{equation}
      \vB = \jaci q R^2 \gz + \gz \times \gr.
\end{equation}
Because of the orthogonality of the $\zeta$ surfaces and $\psi$ surfaces in this
coordinate system, the tangential derivative is the same such that we can write
\begin{equation}
      \vB = F \gz + \pg \gct.
\end{equation}
Equating the two expressions for the covariant components allows us to write
the familiar form for the symmetric angle~\cite{Freidberg,HazeltineMeiss},
\begin{equation}
      \vB = F \gz + \gz \times \gr,
\end{equation}
with the Jacobian in this system given as
\begin{equation}
      \jac = \frac{q R^2}{F}.
\end{equation}
This is sometimes referred to as the PEST angle because the first paper describing it
was in a paper by Grimm, Greene, and Johnson~\cite{Grimm76} describing the PEST
ideal MHD code. 
    \newpage \clearpage
\section{Geometric Quantities and Frenet's Local Frame of Reference}

Expressions using the quantities of the first section such as $\sigma$ or $\Pp$,
or the metric elements of the flux coordinate system, can be difficult to use
for gaining physical intuition.  In this section, common geometric quantities
are introduced and related to the reference frames of the previous sections.

Considerations of a particle tracing out a curve in three-dimensional Euclidean
space lead to studies in the mid-19th century of the geometry of curves.  This
led to the development of the Frenet frame of reference, which uses as its local
basis vectors the tangent of the curve, the curvature vector which is normal to
the tangent, and the binormal vector.  The fundamental theorem of space curves
states that in three-dimensional space every curve with non-zero curvature has
its shape completely determined by curvature and torsion.  Because of the
importance of the $BC$ frame of reference, the relationship of this frame to the
curvature and torsion has been considered for some time~\cite{Dewar84,Hegna00}.
That is, while the curvature and torsion of $\vJ$ might also be of interest, we
will only consider the curvature and torsion of the magnetic field. Our goal is
to express the Frenet quantities of torsion and curvature in terms of our
quantities from the $BC$ and $JK$ reference frames.  

Hegna~\cite{Hegna00} related the quantities of torsion to the normalized version
of our $BC$ frame (also known as the Stix frame).  The relationships are:
\begin{eqaligned}
	  \lp \bhat \cdot \grad \rp \normal &=
	  			-\curvn \bhat - \ntorsion \binormal
									\nonumber \\
	  \lp \bhat \cdot \grad \rp \bhat &=
	  			\curvn \normal - \curvg \binormal
									\nonumber \\
	  \lp \bhat \cdot \grad \rp \binormal &=
	  			\ntorsion \normal + \curvg \bhat
	  \label{Frenet}
\end{eqaligned}
The curvature, rather than acting as the normal component in the Frenet frame,
is now decomposed into the normal curvature, $\curvn$ that is perpendicular to a
flux surface, and the geodesic curvature, $\curvg$, that is parallel to the
magnetic binormal direction.  In these formulas, the only component of
torsion that matters is the normal torsion, $\ntorsion$.  Here, we briefly
review the relationship of these quantities to our frames, and derive other
important geometric relationships.

\subsection{Magnetic Shear}
Although a single curve can be described by curvature and torsion, it does not
convey how neighboring curves behave.  In magnetically confined systems, how a
magnetic field line moves relative to a neighboring field line is important for
modes that have finite width.   The most useful measure of this variation is the
magnetic shear, $\shear$, which is defined as
\cite{Greene81,JohnsonGreene67}:
\begin{equation}
	  \shear \equiv -\frac{1}{C^2} \vC \cdot \grad \times \vC.
   \label{Shear}
\end{equation}
Definition of local shear often differs from this definition by either a minus
sign \cite{Hegna00} or the factor of $C^2$ \cite{Greene81,JohnsonGreene67}. 

Although the $\vC$ component is generally the most useful component of
$\curl{\vC}$, we will consider the other components as well.  The normal
component can be shown to be zero using the divergence-free nature of
$\vB$:
\begin{equation}
	  \grad \cdot \vB = \grad \cdot \lp \vC \times \vR \rp
	                  = \vR \cdot \grad \times \vC = 0.
\label{CurlCIdentity}
\end{equation}
Using the relation between $\vC$ and $\A$ 
(Eq.\ \ref{CeqA}), the curl of $\vC$ may be written as
\begin{equation}
		\curl{\vC} 	= \lp \vRsq \vC \cdot \grad \gra \rp \vbb
				- \lp \vB \cdot \grad \gra \rp \vcc
	\label{CurlCcomps}
\end{equation}
Using Eq. (\ref{graEq}), the local magnetic shear then is given by 
\begin{equation}
		C^2 \shear = - \vB \cdot \grad \gra
		      = \jaci \lp q' + \dT{\residual} \rp
	\label{ShearRelation}
\end{equation}
Thus $q'$ is termed the global shear, and $\residual$ is termed the
integrated residual shear because its derivative give the variation 
of the local shear within a flux surface.

Performing a similar calculation for the parallel component of the
curl of $\vC$ allows us to write the total curl as
\begin{equation}
	  \curl{\vC} = -C^2 \lb F \shear \vbb + \shear \vcc \rb
	\label{CurlC}
\end{equation}

\subsection{Torsion}
The normal component of the torsion is given by 
\begin{equation}
	  \ntorsion = \bhat \cdot \grad \frac{\vC}{\mid \vC \mid} \cdot
				\frac{\vR}{\mid \vR \mid} 
			= \frac{1}{B^2} \vB \cdot \grad \vC \cdot \vR.
\end{equation}
Using the identity
\begin{eqaligned}
	  2 \vect{a} \cdot \grad \vect{b} \cdot \vect{c} &= 
		    \vect{a} \cdot \grad \lp \vect{b} \cdot \vect{c} \rp
		  - \vect{b} \cdot \grad \lp \vect{c} \cdot \vect{a} \rp
		  + \vect{c} \cdot \grad \lp \vect{a} \cdot \vect{b} \rp
									\nonumber \\ &&
		  - \vect{a} \times \vect{b} \cdot \grad \times \vect{c}
		  + \vect{b} \times \vect{c} \cdot \grad \times \vect{a}
		  - \vect{c} \times \vect{a} \cdot \grad \times \vect{b}
\end{eqaligned}
the normal torsion can be related to other
geometric quantities
\begin{eqaligned}
	  2 \vB \cdot \grad \vC \cdot \vR &=  
				  \vC \times \vR \cdot \curl{\vB} 
				- \vR \times \vB \cdot \curl{\vC} 
									\nonumber \\
				&= \vB \cdot \vJ - \vRsq \vC \cdot \curl{\vC}
	  \label{torsion1}
\end{eqaligned}
or, relating the normal component of the torsion to the above quantity:
\begin{equation}
	  2 \ntorsion = \sig + \shear.
	  \label{NormTorsion}
\end{equation}
This relationship is discussed in both Ref.~\cite{Greene96} and~\cite{Hegna00}
and can be viewed as a type of Grad-Shafranov equation.

\subsection{Curvature}
The curvature of the magnetic field line is defined as
\begin{equation}
	\curv \equiv \lp \bhat \cdot \grad \rp \bhat.
   \label{CurvDef}
\end{equation}
Using the equilibrium force balance ($\vJ \times \vB = \grad P$)
and Ampere's law ($\vJ = \curl{\vB}$), 
the magnetic curvature can be expressed as
\begin{equation}
	  \curv = \frac{1}{B^2} \grad_\perp \lp P + \frac{B^2}{2} \rp.
    \label{CurvForceBalance}
\end{equation}
Because the curvature vector is perpendicular to the magnetic
field, we can write the curvature vector in the contravariant form:
$\curv = \curvr \vR + \curvc \vC$.
The covariant components are trivially related to the 
normal and geodesic curvatures by normalization constants;
$\curvn = \curvr \mid \grad \RR \mid$, 
$\curvg = \curvc \mid \vC \mid$.
Using Eq.\ \ref{CurvForceBalance}, expressions for the 
covariant components of the curvature may be easily calculated:
\begin{equation}
	  \curvr = \frac{1}{B^2} \dr{} \lp P + \frac{B^2}{2} \rp
		    + \frac{\grt}{\grr}
		    	  \frac{1}{B^2} \dT{} \lp \frac{B^2}{2} \rp
    \label{Curvr}
\end{equation}
\begin{equation}
	  \curvc = \frac{1}{C^2 B^2} 
		    \vC \cdot \gt  \dT{} \lp \frac{B^2}{2} \rp
	     = -F \jaci \frac{1}{B^4} \dT{} \lp \frac{B^2}{2} \rp.
   \label{CurvC}
\end{equation}
This will be used in the next section.

An important relation for the geodesic curvature can be derived using 
$\dive{\vJ}=0$ and \refeqs{JKBC}, \ref{CIdentity}, \ref{CurvC}:
\begin{eqaligned}
	  \dive{\vJ} &= \dive{}\lp \sig \vB - \Pp \vcc \rp
	     = \vB \cdot \grad \sig - 
		    \dive{} \frac{\Pp}{B^2} \vRsq \vC
	     = \vB \cdot \grad \sig + \Pp 2 \curvc
\end{eqaligned}
or,
\begin{equation}
	  2 \curvc = -\vB \cdot \grad \sigp.
   \label{CurvCIdentity}
\end{equation}
Thus we can see that the angle between the $BC$ coordinate system and the
$JK$ coordinate system which was shown to be related to $\sig/\Pp$ is
related to the geodesic curvature.

\subsection{Curl($\vK$)}
Due to its relationship to the local magnetic shear, the curl of $\vC$ is an
important quantity.  The curl of $\vK$ is also an important quantity and here we
examine its components.  Similar to the curl of $\vC$, the curl of $\vK$ can be
shown to be perpendicular to $\vR$ using $\dive{\vJ} = 0$ which allows us to
decompose the curl  using $J$ and $K$ components, or $B$ and $\C$ components.
The $K$ component is the definition of the shear of the current density field.
Experience shows that this is not a very useful quantity, so we opt for
examining the
$B\C$ components of the curl.
Calculating the curl of $\vK$:
\begin{eqaligned}
		\curl{\vK} &= \curl{} \vbb + \curl{} \sigp \vC
		      = \sig \vbb + \frac{\vB \times \grad (P + B^2)}{B^4}
				- \vC \times \grad \sigp + \sigp \curl{\vC}
	\label{CurlK1}
\end{eqaligned}
The parallel component is
\begin{equation}
	\vB \cdot \curl{\vK} = \sig - B^2 \dr{}(\sigp)
			         - \frac{\sig C^2}{\Pp} F \shear 
	\label{CurlKB}
\end{equation}
where we have used \refeq{CurlC}.
The perpendicular component is
\begin{equation}
	\vC \cdot \curl{\vK} = -2 \curvr + \frac{P'}{B^2} 
				  - \frac{\sig C^2}{\Pp} \shear
	\label{CurlKC}
\end{equation}
Using the generalized Grad-Shafranov equation (Eq.\ (\ref{GenGS}))
and equations for the normal curvature (Eq.\ (\ref{Curvr})) 
and shear (\refeqs{ShearRelation} and \refeqa{graEq}),
one can derive
\begin{equation}
	\vC \cdot \curl{\vK} = 
		    \jaci \dr{} \jac + \jaci \dT{}\lp \jac \frac{\grt}{\grr} \rp
	          +\frac{\Fp C^2}{\Pp} \shear 
	\label{CurlKC2}
\end{equation}
The importance of this component is discussed in the derivation of the inner
layer equations.

A partial listing of the expressions in this section are summarized in
Table~\ref{tab:coordinateSummary}.

\begin{table}
\begin{center}
\begin{tabular}{LL}
\toprule
   \multicolumn{2}{c}{\textbf{Summary of equilibrium-based local frames of reference}} 
                                                                  \\ \midrule
   \vR = \frac{\vB \times \vC}{\vCsq}       & \vR = \frac{\vJ \times \vK}{\norm \vKsq}
									          \\  
   \vB = \vC \times \vR                     & \vJ = \norm \vK \times \vR 
									          \\  
   \vC \equiv \frac{\vR \times \vB}{\vRsq}  & \vK \equiv \frac{\vR \times \vJ}{\norm \vRsq} 
									          \\  
   \vCsq = \frac{B^2}{\vRsq}                & \vKsq = \frac{J^2}{P'^2 \vRsq} 
									          \Bstrut \\  
   \vB = \sig \vBsq \vjj + \vkk            & \vJ = \sig \vB - \norm \vcc,
									          \Tstrut \\  
   \vC = \frac{\sig \vCsq}{\norm} \vkk - \norm \vjj  & \vK = \sigp \vC + \vbb
									          \Bstrut \\  
   \vect{f} = f^\psi \vrr + f^B \vbb + f^C \vcc \ \ \ \ \ & = f^\psi \vrr + f^J \vjj + f^K \vkk.
									          \Tstrut \\  
   f^B =  \frac{1}{K^2} (\sig \C^2 f^J + f^K)   & f^J =  \frac{1}{\C^2} (\sig \C^2 f^B - \norm f^\C)
									          \\  
   f^\C = \frac{\norm}{J^2} (\sig  B^2 f^K - \norm f^J)  & f^K = \frac{1}{\norm B^2} (\sig  B^2 f^\C + \norm f^B)
                                                      \\ \bottomrule

\end{tabular}
\end{center}
\begin{center}
\begin{tabular}{LL@{\hskip 20pt}l@{\hskip 12pt}l}
\toprule
   \multicolumn{4}{c}{\textbf{Summary of flux coordinate's relationship to equilibrium-based frames}} 
                                                                  \\ \midrule
    \vB & = \gz \times \gr - q  \gt \times \gr & flux coordinates: & 
									          \\  
	   & = F \gcz + \pg \gct                    &$q=q(\psi) $ &   $\Rightarrow \vB$ is straight
									          \Bstrut \\  
    \vC & = \frac{-1}{\vRsq} \lb F \gz \times \gr + \pg \gt \times \gr \rb 
	    & $\pg \defn \jac B^2 - qF$; 
        & $\pG\fr = \thetaav{\pg} = \Vpo \fluxav{B^2} - qF$ 
									          \Tstrut \\  
	    & = \gcz - q \gct &   Boozer:    $\pg=\pG $ &   $\Rightarrow \vRsq \vC$ is straight
									          \Bstrut \\  
    \vJ & =  -\Fp \gz \times \gr - \ph \gt \times \gr
        & $\ph \defn -\jac \Pp - q\Fp$; 
        & $\pG^\prime\fr = \thetaav{\ph} = -\Vpo \Pp - q\Fp$ 
									          \Tstrut \\  
	    & = \lp \sigma   F - \frac{  \Pp}{C^2} \rp \gcz 
            + \lp \sigma \pg + \frac{q \Pp}{C^2} \rp \gct
        &   Hamada: $\ph=\pG^\prime$  
        &   $\Rightarrow \vJ$ is straight
									          \Bstrut \\  
    \vK & = \frac{-1}{\Pp \vRsq} \lb \lp \sigma   F - \frac{  \Pp}{C^2} \rp \gz \times \gr \right.
									 & &     \Tstrut \\  &
	     \ \ \ \ \left.  + \lp \sigma \pg + q \frac{\Pp}{C^2} \rp \gt \times \gr \rb & &
									          \\  
	    & = \frac{-1}{\Pp} \lb \Fp \gcz  + \ph  \gct \rb &  &
                                                      \Bstrut \\ \midrule
   \multicolumn{4}{c}{\textbf{Summary of helical angle coordinate's relationship to equilibrium-based frames}} 
                                                      \\ \midrule
    \vB & = \ga \times \gr &  & 
									          \\  
	   & = F \gca + \pd \gct & &  
									          \Bstrut \\  
    \vC & = \frac{-1}{\vRsq} \lb F \ga \times \gr - \pd \gt \times \gr \rb 
         & $\pd \defn \jac B^2$;    
         & $\pD\fr = \thetaav{\pd} = \Vpo \fluxav{B^2} $ 
									          \Tstrut \\  
	  & = \gca 
         & Boozer:  $\pd=\pD $         & $\Rightarrow \vRsq \vC$ is straight
									          \Bstrut \\  
   \vJ & = -\Fp \ga \times \gr + \Pp \jac \gt \times \gr & &	
									          \Tstrut \\  
	    & = \lp \sigma   F - \frac{  \Pp}{C^2} \rp \gca + \sigma \pd \gct 
        & Hamada: $\jac = \jac(\psi)$  &   $\Rightarrow \vJ$ is straight
									          \Bstrut \\  
   \vK & = \frac{-1}{\Pp \vRsq} \lb 
		  \lp \sigma   F - \frac{  \Pp}{C^2} \rp \ga \times \gr \right.  & &
									          \Tstrut \\  
       & \ \ \ \ \left.  + \lp \sigma \pd \rp \gt \times \gr \rb & &
									          \\  
	 & = \frac{-1}{\Pp} \lb \Fp \gca  - \Pp \jac \gct \rb &  &
                                                      \\ \bottomrule
\end{tabular}
\end{center}
\begin{center}
\begin{tabular}{LL}
\toprule
   \multicolumn{2}{c}{\textbf{Brief summary of key geometric quantities}} 
                                                                  \\ \midrule
   \shear \equiv -\frac{1}{C^2} \vC \cdot \grad \times \vC  &
	  \curl{\vC} = -C^2 \lb F \shear \vbb + \shear \vcc \rb
									          \\  
   \ntorsion \equiv \frac{1}{B^2} \vB \cdot \grad \vC \cdot \vR  &
	  2 \ntorsion = \sig + \shear.
									          \\  
   \curv \equiv \lp \bhat \cdot \grad \rp \bhat   &
	  \curv = \frac{1}{B^2} \grad_\perp \lp P + \frac{B^2}{2} \rp.
									          \\  
	\vC \cdot \curl{\vK} = -2 \curvr + \frac{P'}{B^2} - \frac{\sig C^2}{\Pp} \shear &
                                                      \\ \bottomrule

\end{tabular}
\end{center}
\caption{Summary of the flux and helical-angle coordinate systems with
demonstration of the special cases.}
\label{tab:coordinateSummary}
\end{table}

    \clearpage \newpage
\section{Brief Example of equilibrium-based coordinate calculations: Ideal MHD} 
\label{IdealMHD}


Before discussing the derivation of the inner layer equations, we give
a brief example of how the new formalism simplifies computations that arise
in MHD.  The energy principle is given by~\cite{Bernstein58,Freidberg}:
\begin{equation}
	W_F = \frac{\mu_0}{2} \int dV \lb 
			\lb
				\frac{\vec{Q}}{\mu_0} + 
				\lp \vJ \times \normal \rp \lp \vxi \cdot \normal \rp
			\rb^2
			+ \Gamma P \lp \dive{\vxi} \rp^2
			- 2 \lb \vB \cdot \grad \normal \cdot \vJ \times \normal \rb 
			    \lp \vxi \cdot \normal \rp^2 
		\rb
		\label{DeltaW}
\end{equation}
Expressing this in terms of our equilibrium-based coordinate systems
shows that the vector $\vK$, which has not been discussed before,
features prominently: 
\begin{equation}
	W_F = \frac{\mu_0}{2} \int dV \lb 
			\lb
				\frac{\vec{Q}}{\mu_0} + 
				\frac{\xr}{P'} \vK 
			\rb^2
			+ \Gamma P \lp \dive{\vxi} \rp^2
			+ \lb 2 P' B \mid \grad \RR \mid \bhat \cdot \grad \normal \cdot \vK \rb 
            \lp\xr\rp^2 
		\rb
		\label{DeltaWwK}
\end{equation}
The fact that $\vK$ appears in every both the non-compressive stabilizing term, 
and the de-stabilizing term indicates that this vector is important in MHD
studies.

The factor in the brackets of the second term involving $\vK$ can be simplified
using \refeqs{JKBC} and \refeqa{Frenet}:
\begin{eqaligned}
	2 P' B \mid \grad \RR \mid \bhat \cdot \grad \normal \cdot \vK
		   &= 2 \vRsq C \bhat \cdot \grad \normal \cdot 
           \lb \sigp \vC + \frac{\vB}{\vBsq} \rb
									  \nonumber \\ 
		   &= -2 P' B^2 \lb \sigp \ntorsion + \frac{1}{BC} \curvn \rb
									  \nonumber \\ 
		   &= -2 P'  \curvn \mid \grad \RR \mid - 2 \ntorsion \sig B^2
\end{eqaligned}
which gives a nice expression for the destabilizing term in terms of geometric
quantities.
Using \refeq{NormTorsion}, the torsion can be related to the parallel
current and the shear as well.  These give two different forms of $W_F$:
\begin{align}
 W_F &= \frac{\mu_0}{2} \int dV \lb 
		\lb \frac{\vec{Q}}{\mu_0} + \frac{\xr}{P'} \vK \rb^2
			+ \Gamma P \lp \dive{\vxi} \rp^2
            - \lb 2 P'  \curvn \mid \grad \RR \mid + 2 \ntorsion \sig B^2 \rb 
		\rb
                                    \label{newDeltaW} \\
     &= \frac{\mu_0}{2} \int dV \lb 
		\lb \frac{\vec{Q}}{\mu_0} + \frac{\xr}{P'} \vK \rb^2
			+ \Gamma P \lp \dive{\vxi} \rp^2
            - \lb 2 P'  \curvn \mid \grad \RR \mid + \sig B^2 \shear - \sig^2 B^2 \rb 
		\rb
                                     \label{newDeltaW2}
\end{align}
The first form was first given in DMS~\cite{Dewar84}.  Both forms are presented
in Greene~\cite{Greene96} along with a discussion of their advantages.  One can
see that $\vK$ arises naturally in MHD although it has not formally been
discussed in that way.  By making its use explicit and clarifying its
relationships, derivations involving its use are easier.
\section{Derivation of Inner Layer Equations} 
\label{InnerLayerEquations}


Although fusion plasmas are very nearly ideal, resistivity can still change the
topology near surfaces where the magnetic winding number is rational.  Just as
the flow of wind over an airplane wing, which may be analyzed using boundary
layer theory, resistive modes can be analyzed using boundary layer theory when
the boundary layers occur near rational surfaces.  In this theory, the plasma is
analyzed in two regions: an ``outer region" where the plasma is ideal, and an
``inner layer" where dissipation is important.

In this section, linear equations are found for the inner resistive layer based
on the narrow-layer-width approximation.	The inner-layer equations in
cylindrical geometry were originally derived by Coppi, Greene, and Johnson
(CGJ).  In toroidal geometry, the derivation is much more difficult because the
$1/R$ dependence of the magnetic field causes the poloidal harmonics to be
coupled.  The formalism used in references \cite{JohnsonGreene67,GGJ} is used in
deriving these equations in toroidal geometry, which are presented in the
Glasser-Greene-Johnson (GGJ) paper \cite{GGJ}.  Here we show how the knowledge
of the coordinate systems allows one to more easily derive these relations.

\subsection{Resistive instability ordering}
We will consider a narrow layer width at the $q_s=M/N$ rational surface
where we adopt the usual resistive layer ordering \cite{JohnsonGreene67,GGJ}
\begin{equation} 
	\label{CResOrdering}
	x \equiv \psi - \psi_s \sim \gamma \sim \eta^{1/3} \sim \epsilon,\
	\ \ \dpsi{} \sim \frac{1}{\epsilon}.
\end{equation} 
Equilibrium quantities will be considered to be
approximately constant across the layer.  

We begin with the linearized version of the MHD equations.  Our notation
is such that equilibrium quantities are shown with capital letters, and
perturbed quantities are denoted with lower case letters.  That is,
for an arbitrary quantity $Q$, we write
\begin{equation}
	  Q = Q + \eps q_\pone + \eps^2 q_\ptwo + ...
\end{equation}
The linearized MHD equations are then:
\begin{align}
  \label{PertMom}
	\gamma^2 \rho \xi &= (\curl{\vb}) \times \vB + \vJ \times \vb - \grad p
								\\
  \label{PertOhms}
	\vb &= \frac{\eta}{\gamma} \grad \times \lp \grad \times \vb \rp
		    + \grad \times \lp \xi \times \vb \rp
								\\
  \label{PertPres}
	p &=  \xi \cdot \grad P + \specheat P \lp \grad \cdot \xi \rp
\end{align}
where $\specheat$ is the ratio of specific heats, $\xi$ is the velocity divided
by complex growth rate $\gamma$.  

Because we are considering perturbed quantities near the rational surface,
the $\R, \T, \U$ coordinate system is the most useful because it allows
easy identification of
resonant perturbations when Fourier expanding the perturbed quantities:
\begin{eqaligned}
	f &= \sum_{m,n} f_{m,n}\lp \R \rp e^{i \lp m \T - n \zeta
	\rp} e^{i \gamma t}  
									  \nonumber \\
	    &= \underbrace{f_{M,N} \lp \R \rp e^{-iN\U} e^{i \gamma t}
	                    }_{resonant\ perturbation}
		+\underbrace{ \sum_{m,n \not= M,N} f_{m,n}\lp \R\rp 
					  e^{-i n \U} e^{i \gamma t}
					  e^{i \lp m - n q_s \rp \T }
		    	      }_{non-resonant\ perturbation}  
									  \nonumber \\
	    &= \sum_n f_n\lp \R, \T \rp e^{i \gamma t - i N \U}.
\end{eqaligned} 
where again $q_s = M/N$ is the rational surface of interest.
The last line is the procedure we will adopt for simplicity.  If
after transforming, the Fourier transform of a perturbed quantity,
$f_n$, is independent of $\T$, then only the resonant harmonic is
present.  

The helical coordinate also facilitates the representation of the 
parallel gradient operator:
\begin{eqaligned}
   \label{bdotgrad}
	\vB \cdot \grad f &= \jaci \lp \dT{f} + (q - q_s) \dU{f} \rp
								\nonumber \\
			  &\approx \jaci \dT{f} - \eps iN \lambda x f
\end{eqaligned} 
where 
$\lambda = \jaci q^\prime$ and $x= \R - \R_s$.
In the last line, 
$q-q_s$ was Taylor expanded about the rational surface after
Fourier expanding to give  the approximation
$q-q_s \approx \eps q' x$.
Other directional derivatives may be easily computed in a similar manner
\begin{equation}
   \label{jdotgrad}
	  \vJ \cdot \grad f \approx \Pp iN f -F' \jaci \dT{f} - \eps F' iN \lambda x f
\end{equation} 
\begin{equation}
   \label{cdotgrad}
  \vRsq \vC \cdot \grad f \approx B^2 iN f -F  \jaci \dT{f} - \eps F  iN \lambda x f
\end{equation} 
The fact that straight-field line coordinate systems give convenient
forms for $\vB \cdot \grad$ operators has long been recognized.   This
shows that $\vC \cdot \grad$ and $\vJ \cdot \grad$ are greatly
simplified as well.

As is obvious from the above relations, terms of the form
$\jaci \partial Q/\partial \T$ will often arise.
It will be found useful to eliminate these terms which arise due to non-resonant
harmonics to reduce the problem from a two-dimensional problem in $\R,\T$ to a
one-dimensional problem in $\R$.  To do so, we introduce the averaging operator:
\begin{equation}
	\la Q \ra = \frac{2 \pi}{V^\prime} \oint Q \jac d\Theta,
\end{equation}
If $Q$ is symmetric such that there is no $\Z$ dependence, 
this average is the same as the flux surface average given in \refeq{FSA}.

For the vector quantities, decomposition
using the covariant basis vectors is the most convenient.  
The perturbed quantities are decomposed and ordered as
\begin{align}
  \label{bpert}
	\vb &= \br_\ptwo \vrr + \bb_\pone \vbb + \bc_\pone \vcc
								\nonumber \\
	    &= \br_\ptwo \vrr + \bj_\pone \vjj + \bk_\pone \vkk
								\\
  \label{xpert}
	\vxi &= \xr_\ptwo \vrr + \xb_\pone \vbb + \xc_\pone \vcc
								\nonumber \\
	    &= \xr_\ptwo \vrr + \xj_\pone \vjj + \xk_\pone \vkk
								\\
  \label{ppert}
	p &= p_\pone.
\end{align}
The $\R$ components of the vector quantities are lower order to help 
satisfy the divergence criterion as shown below.  
The ordering also arises from 
viewing the vectors as arising from a potential of order $\eps^2$, and 
the non-$\psi$ components are of lower order because of the $d/d\R$ 
derivative which arises in those components \cite{Kruger99}.

Because the perturbed magnetic field is divergence-free 
($\dive{\vb}=0$), the components can be related to each other:
\begin{eqaligned}
	\grad \cdot \vb 
			    &= \jaci \lb 
			        \dr{} \lp \jac \br \rp
				+ \dT{} \lp \frac{1}{B^2} \lp \bb -F \bc \rp \rp
				+ \dU{} \lp \jac \bc \rp
				\rb + \mathcal{O}(\eps^2)
\end{eqaligned}
Taking the average of the last expression, we have
\begin{equation}
	\dr{} \abr = iN \abc.
   \label{brc}
\end{equation}
The orderings shown in Eq. \ref{bpert} are motivated by this balance.
Looking at the pressure equation, Eq.\ (\ref{ppert}),
we see that the first two terms are of order $\eps^2$ which implies that the
the plasma is incompressible to lowest order, $\grad \cdot \vxi \approx 0$.
An equation similar to Eq. \ref{brc} for the displacement vector can then
be derived.

\subsection{First-order equations and magnetosonic waves}
Our goal is to obtain equations for averaged perturbed quantities
$\abr, \axr, \abb$ by taking the appropriate projections of
the linearized MHD equations and ordering.  The details of this
calculation are presented in the appendix.  Here we summarize the
results of the lowest order equations which gives the components
whose resonant harmonics are dominant at the rational surface:
\begin{align}
	 \vR \cdot {\rm (Momentum\ Eq.)} &\rightarrow
	 p = -\bb
								\\
	 \vR \cdot {\rm (Induction\ Eq.)} &\rightarrow
	  \fluxav{\xr} = \xr
								\\
	 \vB \cdot {\rm (Momentum\ Eq.)} &\rightarrow
	  \fluxav{\bb} = \bb
   \label{bbeq0}								 
								\\
	 \vJ \cdot {\rm (Momentum\ Eq.)} &\rightarrow
	  \fluxav{\bj} = \bj
   \label{bjeqz}						\\
	 \vC \cdot {\rm (Induction\ Eq.)} &\rightarrow
	  \fluxav{\xc} = \xc
   \label{xceq0}								 
								\\
	 \vK \cdot {\rm (Induction\ Eq.)} &\rightarrow
	  \fluxav{\xk} = \xk
   \label{xkeq0}								 
\end{align}
The first equation is the equilibration of compressional Alfv\`en waves to
lowest order.  The second, third, and fifth equations are
the equilibration of sound waves to lowest order.  The fourth equation
is a statement of $\dive{\vJ}=0$ lowest order, and the final equation is
a statement of $\dive{\vV}=0$ to lowest order.
Using Eq. \ref{JKcomps} and $\abj = \bj$, we can find the
variation of $\bc$ within a surface:
\begin{equation}
	  \bc = \frac{\C^2}{\acsq} \abc + \frac{1}{\Pp} \bb 
		    \lp \sig \C^2 - \ascsq \frac{\C^2}{\acsq} \rp.
   \label{bcwithinsurface}
\end{equation}
Similarly, using Eq. \ref{JKcomps} and $\axk=\xk$, we can find the
variation of $\xb$ within a surface:
\begin{equation}
	  \xb = \frac{B^2}{\absq} \axb - \frac{1}{\Pp} \xc 
		    \lp \sig B^2 - \asbsq \frac{B^2}{\absq} \rp.
   \label{xbwithinsurface}
\end{equation}
One can find similar relations for $\bk$ and $\xj$ using 
\refeqs{BCcomps}, \ref{bbeq0}, and \ref{xceq0},
but they are not needed for this derivation. 

\subsection{Annihilation operators}
We are now ready to look at the next order of the equations.  
The $\vC$ component of the momentum equation will yield the same 
information as the $\vR$ component of the magnetic field to
lowest order; i.e., that magnetosonic waves need to be eliminated
(Eq.\ \ref{magnetosonic}).  To obtain additional information from
this equation, we must go to higher order and subtract the two components
from each other.  A convenient way to do this is to find an operator
that annihilates the lowest order information, apply it to the full
equation before ordering, and then take the lowest-order terms.
This is the concept of annihilation introduced by Kruskal.
CGJ \cite{CGJ} and GGJ \cite{GGJ} used as the annihilation operator
\begin{equation}
	\grad \cdot \vbb \times
\end{equation}
This operator was so important to MHD theory that the Princeton theory group
dubbed it the Grand Old Operator~\cite{coppi-private}.
In toroidal geometry and considering the effects of compressibility\cite{GGJ}, 
this operator leaves a higher order term, $\xi^\C_\pthree$.  To cancel
this term, the GGJ equation is formed by
\begin{equation}
	\left< \grad \cdot \vbb \times (Momentum\ Eq.) 
	- \sigp \dr{}\vB \cdot(Momentum\ Eq.) \right>
\end{equation}
Comparing the equation for the complete formation of the annihilated
equation to the relation of $\vK$ to $\vB$ and $\vC$ (Eq.
(\ref{JKBC}))motivates a new annihilation operator:
\begin{equation}
		\grad \cdot \vK \times
\end{equation}
We study the derivation of the annihilated equation in detail to show
its advantages.

Applying this operator to the momentum equation, we have
\begin{equation}
	\grad \cdot \vK \times \lb \gamma^2 \rho \xi 
      = (\curl{\vb}) \times \vB + \vJ \times \vb - \grad p \rb.
\end{equation}
The first term is
\begin{eqaligned}
	\gamma^2 \rho \grad \cdot \vK \times \xi &= 
	\gamma^2 \rho \grad \cdot 
            \xr \frac{\vJ}{\Pp \vRsq} - \xj \frac{\vR}{\Pp \vRsq} 
                                               =
	\frac{-\gamma^2 \rho}{\Pp} \dr{\xj} + \order{2}.
\end{eqaligned}
The second term is
\begin{eqaligned}
	\grad \cdot \vK \times (\curl{\vb}) \times \vB &=
	     \grad \cdot \lp \vK \cdot \vB (\curl{\vb})\rp 
	    -\grad \cdot \lp \vK \cdot (\curl{\vb}) \vB \rp 
								\nonumber \\
          &=
	   \vB \cdot \grad \lp 
                  \dive{\vK \times \vb} 
			-\vb \cdot \curl{\vK} \rp
								\\
          &=
	   \vB \cdot \grad \lb 
                  \dive{\lp \br \frac{\vJ}{\Pp \vRsq} 
                          - \bj \frac{\vR}{\Pp \vRsq} \rp}
			-\vb \cdot \curl{\vK} \rb
								\nonumber \\
          &=
          iN \lambda x \frac{1}{\Pp} \dr{\bj} 
          -\jaci \dT{} \lp ... \rp 
                           + \order{2}.
								\nonumber \\
\end{eqaligned}
The cancellation used in going from the first step to the second
required $\vK \cdot \vB = 1$.  This is the elimination of the
magnetosonic wave required of our annihilation operator.

The third term is
\begin{eqaligned}
	\grad \cdot \vK \times \lp \vJ \times \vb \rp &= \dive{\bk \vJ}
					= \vJ \cdot \grad \bk
            = 
		-F^\prime \jaci \dT{\bk} + \Pp iN \bk  + \order{2}.
\end{eqaligned}
The fourth term is
\begin{eqaligned}
	\grad \cdot \vK \times \grad p &= \grad p \cdot \curl{\vK}
					= -\dU{\bb} \gradt \U \cdot \curl{\vK}
              = 
		  	\EC \ iN \bb + \order{2}.
\end{eqaligned}
Putting all of the terms together and taking the average, we have
\begin{equation}
	\frac{\gamma^2 \rho}{\Pp} \dr{}\axj +iN\Pp \abk 
                        +iN \fluxav{\EC} \bb
                        +iN \Lambda x \frac{1}{\Pp} \dr{\bj}=0.
   \label{IL3}
\end{equation}
This is the annihilated momentum equation.  All that remains is to
convert the above variables to the variables we are solving for
$\abr, \axr, \abb$ using (\refeqs{BCcomps}-\ref{JKcomps}) and our
previous relationships.

\subsection{The inner layer equations}
Similar to the annihilated momentum equation, the parallel induction
equation leaves higher order terms in toroidal geometry.  The best annhilation
operator for this term is $\vK \cdot$ for this equation.  Applying it yields
\begin{equation}
	\gamma \bk = \eta \drs{\bk} \vRsq 
		    - \dive{\vxi} + \vB \cdot \grad \xk + \EC \xr
\end{equation}
Taking the average and substituting in the expression for $\dive{\vxi}$
from the pressure equation yields
\begin{equation}
	\abk = \eog \la \drs{}\bk \vRsq \ra 
		  - iN \Lambda x \xk
		  + \frac{\bb}{\gamma \specheat P }
		  + \frac{\xr}{\gamma} \lp\fluxav{\EC}+\frac{\Pp}{\specheat P}\rp
   \label{IL4}
\end{equation}
Again, we need to convert the variables in the above equation.  After a
bit of algebra (see Appendix), the normalized inner layer equations
are:
\begin{align}
	   &\Sv_{XX} - H \Bv_X = Q \lp \Sv - \X \Xv \rp,
								\\  &
	     \Gv_{X} = H \Sv_{XX} + F \Bv_X,
								\\  &
	     Q^2 \Xv_{XX} - \X^2 Q \Xv + E \Bv + Q \X \Sv + \Gv = 0,
   \label{NormAnnhilatedMom}
								\\  
        & \frac{1}{Q^2} \Bv_{XX} - \frac{\X^2}{Q^2} \Bv + G \Bv 
	     + \frac{\X}{Q^2} \Sv - K \Gv + (G-KE)\Xv = 0.
   \label{NormAnnhilatedInduction}
\end{align}
\noindent
where
\begin{eqaligned}
	   &
	   	E = \CE
								\\  &
	   	F = \CF
								\\  &
	   	H = \CH
								\\  &
	     G = \CG; \ \ \ \ \ \ \ \ \ 
         K_{GGJ} = \CK
								\\  &
	     M = \CM
\end{eqaligned}
\noindent
\refeqs{NormAnnhilatedMom} and \refeqa{NormAnnhilatedInduction} are the
normalized annihilated momentum and induction equations.  As can be seen by the
complicated factors in $F$ and $H$, many of the complications appearing in going
from \refeqs{IL3} and \refeqa{IL4} to these equations are due to converting the
$J$ and $K$ contravariant components which arise naturally into the normal and
parallel components.  Examination of \refeqs{BCcomps} and \refeqa{JKcomps} shows
easily how many of the factors arise.

In GGJ, it shown that a plasma is ideal unstable if $D_I>0$ where
$D_I=E+F+H$
and that it is unstable to resistive interchange modes if
$D_R=E+F+H > 0$.
From the expression for $E$ given above it is not obvious that our form
reduces to forms given previously in the literature.  We now show that it does.

Using the equation for $\EC$ (Eq.\ (\ref{CurlKC2})) and the above
relation, one can write
\begin{equation}
	   \EC = \jaci \dr{} \jac +\frac{F^\prime}{\Pp} C^2 \shear
	               + \jaci \dT{}\lp \jac \frac{\grt}{\grr} \rp.
\end{equation}
Taking the average of this equation and substituting into the
equation for $E$, yields 
\begin{equation}
	  E = -\frac{\la C^2 \ra \Pp}{\Lambda^2} \lb
	     \frac{\Vpp}{\Vp} + \frac{4 \pi^2 F^\prime q^\prime}{\Vp\Pp} 
	     + \frac{\Lambda}{\Pp} \frac{\la \sigma B^2 \ra}{\la B^2 \ra} 
	     \rb
\end{equation}
which gives the more familiar $V^{\prime\prime}$ criterion of GGJ
\cite{GGJ,Greene97}.  The direct calculation of $E$ is
easier in Hamada coordinates because in calculating the curl of $\vK$,
it is easier when $\vJ$ is straight and $\vK$ has
a nice contravariant form seen in Eq.\ (\ref{K_contra}).
\section{Discussion and Summary} 
\label{Summary}

Two equilibrium-based local frame of references have been introduced.
The $BC$ reference frame is widely used even if not explicitly stated as such.
When normalized, it is closely related to the explicitly named Stix frame (differing
only in that the parallel direction is last and the binormal changed
accordingly)~\cite{wright_thesis1998}, and to the Frenet-like frame discussed by
Hegna~\cite{Hegna00}.  Decomposition of fields into the $BC$ local frame was
explicitly performed by MHD theory in the 60's~\cite{CGJ}.  The usefulness of
having $\vC$ straight was pointed out by Boozer in the early
80's~\cite{Boozer81,Boozer82}.

It was DMS~\cite{Dewar84} that made the transformation between flux coordinates
and the local $BC$ frame explicit.  This systematic investigation is useful for
the algebraic expressions that consistently arise in analytic derivations.
The original motivation of the DMS paper is in discussing numerical
implementations.  Because different analytic forms of the linearized MHD
equations can have different convergence properties, being able to analytically
transform easily is important.  

An equivalent systematic investigation of the transformation properties of the
$JK$ frame has never been performed, and is new to this paper.  The fusion theory
community has developed a battery of techniques, both analytic and
computational, for dealing with the stiff time scales of plasma oscillations,
compressional Alfv\`en waves, and (not discussed in this paper) cyclotron motion.
One of the main methods for dealing with compressional Alfv\`en waves
analytically, the Grand Old Operator, actually does not quite work in toroidal
geometry.  It is seen that $\dive{\vK \times}$ is the appropriate annihilation
operator, and that the dominant term of $D_I$ appears naturally.  The other
terms in $D_I$ in this derivation are the metric elements of the $BC$ and $JK$
frames that arise when converting the annihilated equation variables to the
fundamental variables that we wish to solve for.  By having a clean separation
from the two frames, the derivation is clearer.  The magnetic well term, as
embodied in the $V''$ criterion, is shown to be related to the magnetic binormal
direction of the curl of the current density binormal.  The important role of
$\vK$ perhaps should have been obvious sooner given that the ideal MHD potential
energy term, $W_F$, is nicely expressed in terms of $\vK$ and the manipulations
thereof.

The presentation here focused on the calculation of $W_F$ and the inner layer
equations, but the formulas and techniques presented here should be useful for
any theoretical analysis in toroidal geometry where one wishes to separate the
length scales associated with different directions.

\begin{acknowledgments} 

This manuscript would not exist without the ideas and insights of Dr. John M.
Greene.  Dr.~Greene asked the primary author for aid in expanding these ideas
after helping him with his thesis work.  It is unfortunate that the fulfillment
of these ideas occurred after illness and subsequent death prevented him from
providing further inspiration.

The primary author would also like to thank Drs. Chris Hegna, James Callen, and
Alan Turnbull for useful discussions, and Dr. Tom Jenkins for reviewing the
manuscript.  This material is based on work supported by US Department of
Energy, Office of Science, Office of Fusion Energy Sciences under award numbers
DE-SC0019067.

\end{acknowledgments}

	\bibliographystyle{apsrev}
	\bibliography{merged}

	\appendix
	\include{details}
\end{document}